\documentclass[11pt]{article}
\usepackage[utf8]{inputenc}
\usepackage{lmodern} 
\usepackage[english]{babel}
\usepackage[autostyle]{csquotes}
\usepackage[backend=bibtex8,style=authoryear,doi=false,isbn=false,url=true,uniquename=false, giveninits=true, natbib]{biblatex}
\usepackage{csquotes}
\usepackage{dcolumn}
\usepackage{tabularx, ragged2e}
\usepackage[tight-spacing=true]{siunitx}
\usepackage{authblk}
\usepackage[a4paper,top=3cm,bottom=2cm,left=3cm,right=3cm,marginparwidth=1.75cm]{geometry}
\usepackage{ntheorem}
\newtheorem{hyp}{Hypothesis}
\usepackage{array}
\usepackage[super]{nth}
\usepackage{tabularx}

\usepackage{amsmath}
\usepackage{graphicx}

\AtEveryBibitem{%
  \clearlist{language}%
}

\title{Explaining the emergence of echo chambers on social media: the role of ideology and extremism}

\author[1]{Jonathan Bright \thanks{jonathan.bright@oii.ox.ac.uk}}
\affil[1]{Oxford Internet Institute, University of Oxford}

\begin{document}

\maketitle

\begin{abstract}

The emergence of politically driven divisions in online discussion networks has attracted a wealth of literature, but also one which has thus far been largely limited to single country studies. Hence whilst there is good evidence that these networks do divide and fragment into what are often described as ``echo chambers'', we know little about the factors which might explain this division or make networks more or less fragmented, as studies have been limited to a small number of political groupings with limited possibilities for systematic comparison. 

This paper seeks to remedy this deficit, by providing a systematic large scale study of fragmentation on Twitter which considers discussion networks surrounding 90 different political parties in 23 different countries. It shows that political party groupings which are further apart in ideological terms interact less, and that individuals and parties which sit at the extreme ends of the ideological scale are particularly likely to form echo chambers. Indeed, exchanges between centrist parties who sit on different sides of the left-right divide are more likely than communication between centrist and extremist parties who are, notionally, from the same ideological wing. In light of the results, theory about exposure to different ideological viewpoints online is discussed and enhanced. 

\end{abstract}

\section{Introduction}

The online environment offers an almost infinitely wide choice when it comes to what type of information to consume and what type of people to converse with; changing between different sites and channels is, furthermore, straightforward \citep{Prior2005}. Hence online environments are structured by the choices of the people operating in them (unlike their offline counterparts where significant constraints operate, for example a limited number of newspapers or television channels, or a geographically constrained social circle of people to talk with). 

The fluidity of these environments has stimulated a great deal of research on the types of social structure which end up developing online, particularly when discussions about politics are involved. A key strand of this research has concerned what could be referred to as the ``echo chamber'' thesis: the idea that online conversations about politics are typically divided into a variety of sub-groups, and that this division takes place along ideological lines with people only talking to others with which they are already in agreement. Concerns about political division and fragmentation online were voiced in some of the earliest theoretical work on the Internet \citep{Dahlberg2007,Papacharissi2002,Sunstein2009,VanAlstyne1999}. Numerous empirical studies in a variety of Internet discussion contexts such as forums \citep{Edwards2013,Hill1998}, blogs \citep{Adamic2005,Hargittai2008,Lawrence2010} and most recently on social media \citep{Aragon2013,Barbera2014,Colleoni2014,Conover2011,Conover2012,Feller2011,Gaines2009,Garcia2015,Gruzd2014,Himelboim2013,Yardi2010,Quattrociocchi}\footnote{However many studies have also shown at least some evidence of cross ideological exposure \citep{Wojcieszak2009OnlineDisagreement}, whilst others have argued that the Internet makes a positive overall contribution to the heterogeneity of political networks \citep{Brundidge2010,Kim2011ThePerspectives}} have since found evidence that at least some degree of fragmentation exists. These patterns have concerned many theorists of democracy, who have argued that exposure to a diverse range of viewpoints is crucial for developing well informed citizens \citep[p.2]{Gentzkow2010} who are also tolerant of the ideas of others \citep{Nunn1978}. By contrast, exposure to only like-minded voices may contribute towards polarization towards ideological extremes \citep{Sunstein2002,Warner2010}\footnote{Though others have countered that too much cross ideological exposure can lead to political ambivalence, and hence at least a certain amount of fragmentation is necessary for political action \citep{Dahlberg2007,Mutz2002}}. As social media networks become more important for shaping political viewpoints and exposing people to information \citep{Bakshy2015,Bright2016}, the social relevance of the echo chamber thesis is only likely to grow. 

However, while the literature on the subject is rich, empirical studies which focus on the structure of networks as a whole (and the extent to which echo chambers form within them) have thus far have been largely descriptive (although much analytical research exists at the individual level, which is treated more fully below). This means that, though these studies (albeit using different methodologies) often appear to show different absolute levels of fragmentation between individual groups in the networks they study, we know little about the mechanisms which might drive this variation. The major reason for this is that thus far the vast majority of studies which have looked at the network level have been single country ones, with an overwhelming focus on the US, as Garcia et al. have remarked \citep[p.46]{Garcia2015}. This provides little scope for studying variation (as each single country network typically contains only a handful of groups), and hence potentially explaining some of the factors which might lead to more or less fragmentation. The aim of this study is to take a first step towards remedying this deficit, by seeking to explain variation in levels of political fragmentation through a systematic large scale study. The research question is simple: what explains fragmentation in online discussion networks? 

The rest of the article is structured in the following way. Section one offers a clearer definition and conceptualization of the ideas of fragmentation and echo chambers, and builds up theory and hypotheses about why they might both emerge and vary in intensity, focusing particularly on ideological variations between different sub-groups in the network in question. Section two outlines the methodology employed, explaining the collection of data from Twitter concerning political discussions in a variety of different EU countries during the European Parliament elections in 2014, and describing how key concepts are operationalised. Section three, finally, sets out the results, with evidence showing that ideology does indeed appear to drive variation in the intensity of echo chambers, with the role of ideological extremism being particularly important. The results are then discussed in light of their potential consequences for the circulation of ideas online. 

\section{Explaining Echo Chambers in Online Discussion Networks}

This first section will conceptualise echo chambers in online discussion networks in general, and also explore why ideological factors might be likely to provoke greater or lesser levels of intensity in these echo chambers. In so doing, the section will also pose the hypotheses to be tested. 

For the purposes of this article, online discussion networks are spaces such as forums, blogs and social media sites where people can engage in conversation and exchange messages. In conceptual terms, echo chambers emerge when these networks become fragmented into different groups of people along ideological lines (meaning that within each group individuals are, broadly, in ideological agreement). Groups themselves can be defined through patterns of communication: a group forms in the network if its members communicate with each other relatively frequently compared to their levels of communication with non-group members. If a space is fragmented, and hence contains echo chambers, there will be relatively weak patterns of communication between groups, and relatively strong patterns of communication within groups \citep[p.47]{Garcia2015}. By contrast, a space which is not fragmented will show lively patterns of between group communication; indeed, it may be hard identify or delineate sub-groups at all just by observing communication patterns. 

This definition of suggests that the measurement of echo chambers cannot take place at the group level (i.e. we cannot just measure whether one particular group is an echo chamber) because identification of echo chambers requires measurement of \textit{within} group communications relative to \textit{between} group communications. However, it would also be problematic to try and measure the level of fragmentation at the network level as a whole. When a political discussion network consists of more than two groups (as might be expected, for example, in a political system which has more than two major parties) trying to identify an overall level of fragmentation for the network might disguise important local variations: lively discussions between some of the groups in the network might be counterbalanced by the separation or isolation of other groups from the network. Hence, in this article the focus is placed in explaining fragmentation observed between different \textit{pairs} of groups within a wider discussion network. In addition to making measurement more tractable, focussing on the ``group-dyad'' also allows hypotheses to be developed based on both the individual characteristics of groups in question and the relationship between group characteristics.

Theoretical explanations for the emergence of echo chambers have, thus far, focussed on micro level behavioural characteristics. Fragmentation of political discussion networks is a self-organising phenomenon which emerges from the behavior patterns and choices made by individuals participating in communication, hence the focus on individual level motivations is logical. Several mechanisms have been identified. The most commonly identified cause is the tendency towards ``homophily'' \citep{Barbera2014} an impulse to form social ties with others who are similar to oneself in some way \citep{McPherson2001}. Such a tendency means that ideological fragmentation will naturally emerge in online discussions as people connect to others with similar views. Closely related to homophily is the concept of ``selective exposure'' \citep{Knobloch-Westerwick2009}, a phenomenon whereby people select information or sources they already agree with whilst filtering out others \citep{Garrett2009}. If we consider online discussions as a source of information, then the selective exposure mechanism will produce similar results to the homophily mechanism, as people select themselves into online discussions with which they already agree whilst ignoring others with which they disagree. A further mechanism identified is the tendency of individuals to moderate their opinions into line with what they perceive the group norm to be \citep{Garcia2015}, or at least to keep quiet if they believe themselves to be outside this norm \citep{Scheufle2000}. This mechanism means that existing groups of association should become more homogeneous over time (or at least they will appear to). The potential anonymity offered by Internet mediated communication may further increase this process, as this anonymity encourages a ``de-individuation'' which makes adopting group norms more likely \citep{Spears1990}. 

It is also worth highlighting a strand of research which offers reasons to expect that echo chambers will not emerge, or at least that they will be moderated in their intensity. These works have highlighted other motivations which encourage people to specifically seek out diverging or opposing viewpoints. For example, Garrett has highlighted that even though people may be less likely to seek opinion challenging information, they also may spend more time engaging with it \citep{Garrett2009}. Furthermore, Kim et al. have argued that diversity of information exposure is linked to personality traits, with introverted and less open individuals more likely to be exposed to cross-cutting information \citep{Kim2013InfluenceTraits}. Theories in this area help explain why, in general, echo chambers are often not complete when they are observed in the real world, and that evidence of cross cutting ideological exposure is also common. 

These micro level mechanisms are in general well known and have been widely tested. However, by themselves they offer little reason to expect why levels of observed fragmentation might vary between groups. Such variation could only be caused by factors which themselves vary between groups, and which contribute to either enhancing or moderating the mechanisms which have been identified as creating echo chambers. The most obvious factor which varies between groups is their ideological stance. Hence, in this article, the major theoretical and empirical focused is placed on the role of ideology in driving echo chamber formation.  

Ideology could have several effects on the micro level mechanisms identified above, and hence the extent to which pairs of groups interact. Firstly and most obviously, it is theoretically plausible that as the ideological distance between groups increases, the fragmentation between them also increases. As groups drift further apart in ideological space, members of each group are likely to perceive those from the other group as ideologically less similar, and are hence they are less likely to make connections to them (following the homophily mechanism). They are also likely to disagree with their arguments more, and hence more likely to selectively ignore their communication (following the selective exposure mechanism). This line of theory leads to the first hypothesis tested in this article:

\begin{hyp}
As the ideological distance between groups increases, they will interact less
\end{hyp}

Secondly, it is also possible that the type of ideology has an impact on the tendency of the group to fragment. Groups on the left and right of the political spectrum attract support from different socio-demographic strata; and it is possible that socio-demographic factors have an influence on predisposition towards political fragmentation. This idea is supported by work which has found, for example, that type of political affiliation appears to have an impact on some measurements of selective exposure \citep{Garrett2009a}. It is also supported by descriptive works on political fragmentation which have repeatedly shown that groups on different sides of the left-right political division have different internal patterns of communication, with some more densely interconnected than others. No clear consensus has emerged as to the direction of the relationship: some studies have found the right wing end of the spectrum to be more densely connected \citep{Conover2011,Hargittai2008,Warner2010}, whilst others have found the reverse \citep{Barbera2014}; one study even found evidence for both conclusions using different measures \citep{Colleoni2014}. However, many studies have shown that some variation along left-right lines exists. 

Furthermore, there are also qualitative differences in terms of ideology between the two sides of the left-right spectrum which might further accentuate this trend. Individuals from groups which sit on either side of the division (even if not very far apart in terms of the scale of the axis itself) are nevertheless likely to perceive each other as dissimilar, and hence may have a tendency to talk to each other less than groups which are a similar distance apart but on the same side of the scale \citep[some evidence for this idea is provided in][]{Feller2011}. This leads to the second hypothesis to be tested in the article: 

\begin{hyp}
Groups from different sides of the left-right divide will interact less than groups from the same side
\end{hyp}

Finally, the ``extremism'' of a group's ideology may also play a role. Again, this relates to potential qualitative differences between extremists and centrists, rather than simply ideological distance. As Stroud has shown \citep{Stroud2010}, individuals with attitudes at more extreme ends of the ideological scale show more pronounced tendencies towards selective exposure than those in the middle, a result attributed to the increased certainty these individuals typically have in their beliefs \citep[see also][]{Johnson2009CommunicationBlogs,Wojcieszak2009,Sunstein2009}. Elements of this line of thinking can also be seen in ``hostile media'' theory, where people with strong pre-existing opinions are more likely to perceive alternative viewpoints as biased, and hence ignore or filter them out \citep{Kim2011PublicApproach}. This dynamic could also be self-reinforcing, as discussion with like-minded individuals can also lead to the polarization of attitudes towards ideological extremes, which in turn stimulates further fragmentation \citep{Huckfeldt2004,Myers1975}. This branch of theory leads to the final hypothesis tested in this article:

\begin{hyp}
Pairs of extremist groups will interact less than pairs of centrist groups
\end{hyp}

While this article places a major focus on ideology as a driver for political fragmentation, there are also a variety of other factors which are worth considering as control variables. First, the overall size of a political grouping might make a difference: larger political groupings might be less likely to communicate with smaller ones as they might be perceived as less worthy of consideration. This idea is supported by Aragón et al. who found diverging communication patterns between small and large political groupings \citep{Aragon2013}. Furthermore, the status of different political groups in the wider political system also differs: some will be related to political parties which are incumbent in government at any given time, whilst others may be in opposition. Previous research has shown that incumbent political forces often make less use of online democratic opportunities \citep{Herrnson2007}: it may be that online groups which are related to incumbent political forces are hence also less connected to the rest of the discussion network as a whole. Of course, it should be noted as well that group size and factors such as incumbency are also potentially correlated with ideology (with, for example, groups on the extreme ends of the ideological spectrum perhaps more likely to be smaller). Including these control variables is hence important in that it will allow identification of the impact of ideology independent of other factors such as size.

\section{Methodology}

The main aim of this paper is to collect a sufficiently large sample of pairs of discussion groups within networks, such that the hypotheses identified above about echo chambers can be tested at sufficiently large scale. In this section, the methodology employed to achieve this will be described. 

Data for the study is taken from the social media network Twitter. Twitter is the only major social media network which is both frequently used for political discussion in a wide variety of countries and which makes its data generally available for research purposes, and it has hence been widely used in previous research on echo chambers. The strategy most commonly deployed in the literature to construct a sample of data from Twitter \citep[applied in, for example,][]{Aragon2013,Conover2011,Feller2011,Garcia2012,Himelboim2013} is the collection of Tweets containing politically relevant keywords or hashtags, often around the time of a key moment in national politics such as an election. This data collection allows the observation of communication activity within what have been described as ``ad-hoc publics'' \citep{Bruns2015} which form through the discussion activities of people using these keywords when they address comments at each other. The structure of relations within these discussions can then be regarded as a directed network, with ``nodes'' in the network representing individuals and directed ``edges'' going from one individual to another when one person creates a tweet containing the name of another person. Various techniques can then be used to determine the extent to which this network is fragmented (the measurement of fragmentation is discussed further below). 

This paper follows this research strategy to an extent, however instead of using hashtags it makes use of a list of all official Twitter account names (or ``handles'') of major political parties and party leaders in all 28 EU member states which was collected by the \textit{euandi} project \citep{GARZIA2015} in the run up to the 2014 European Parliament elections. Making use of the European Parliament elections is useful in this case because it allows observation of a relatively large sample of discussion groups in a variety of countries which are nevertheless broadly comparable, as they are all going through the same electoral process at the same time. For each of these accounts, tweets were collected which either were created by these accounts or which mentioned these accounts (including replies and retweets) through the Twitter streaming Application Programming Interface [API], which is a web service which allows structured data collection from the Twitter platform. This allowed observation of mini-publics which formed through discussions around individual politicians and political actors. Only one handle was used per political party, with preference given to the handle of the party leader; if that was not available, the handle of the party itself was used instead. 

\begin{table}[!htbp]
\centering
\caption{Example of link formation in a political discussion network on Twitter. Directed links are formed from the handles in the User column to all the handles in the Tweet column.}
\label{edge-create}
\begin{tabular}{p{0.15\linewidth}>{\raggedright}p{0.4\linewidth}p{0.3\linewidth}}
\hline
\hline
User & Tweet & Edges Created \\ 
\hline
@blueparty & .@stephanie and @emilyw this am delivered a petition calling for prison reform & @blueparty $\rightarrow$ @stephanie @blueparty $\rightarrow$ @emilyw \\
\\
@john &	@blueparty @stephanie @emilyw didn’t we have a vote on this? Less than 6 years ago? And the answer was no? & 	@john $\rightarrow$ @blueparty \newline @john $\rightarrow$ @stephanie \newline @john $\rightarrow$ @emilyw \\
\\
@paul &	@john @blueparty @stephanie @emilyw Nope. We had a vote specifically on reducing sentences not on reform in general &	@paul $\rightarrow$ @john \newline @paul $\rightarrow$ @blueparty \newline @paul $\rightarrow$ @stephanie \newline @paul $\rightarrow$ @emilyw \\

\hline\hline
\end{tabular}
\end{table}

Table \ref{edge-create} provides an illustrative example of the type of conversation networks created by this data collection method. The conversation described in the table is real, though the Twitter handles have been changed and the text altered slightly to preserve the privacy of the participants. Within the conversation network, nine edges are created by different individuals mentioning others in their messages (with the individuals themselves being the nodes in the network). For example, an edge is created going from @paul to @john, as these two individuals discuss the activities of a series of politicians, who themselves were mentioned in a tweet by the official Twitter account of @blueparty. The data collection strategy allowed the observation of all of the edges of this type formed during the period around the discussion of the Twitter handle in question. Furthermore, as some people discussed multiple political accounts during the electoral period, edges are created between different groups discussing different politicians. This allows quantification of the extent to which different groups are connected, or not, to other groups within the rest of the network. 

This Twitter handle focussed data collection strategy was chosen because using account names as a means of data collection instead of hashtags eliminates the potential for country specific bias and makes large scale data collection feasible: hashtags are inevitably language specific, and often country specific (referring to particular events within national political life) and hence choosing equivalent hashtags in different country contexts would be difficult. By contrast, the relevant Twitter handle of a given politician or political party is straightforward to identify. Focussing on handles rather than hashtags also allows discussion groups to be assigned ideology scores and classified along political lines, as each one is of course centred around a given politician or political party (this assignment and classification is discussed further below). 

Whilst this data collection strategy is effective in terms of generating a large sample of groups, its limitations when compared to the hashtag or keyword based approach should however be acknowledged. The data collected represents only a partial account of all the political discussion which took place on Twitter during the moment of the election: tweets which related to the election but which did not mention a party account will not be collected (of course, hashtag or keyword based collection also suffers from this problem, but arguably to a lesser extent). The choice of studying an election period in general and the EU parliament elections in particular is also not unproblematic: these elections are often perceived as ``second order'' elections (i.e. subordinate to national politics rather than focused on issues at the EU level - see \cite{Reif1980}). There is hence an important question concerning whether the patterns found could be generalized to first order national elections, or indeed politics outside of election time. However, both these sacrifices are necessary in terms of allowing a large number of discussion networks to be collected: any attempts to collect complete discussion records or study national elections would have necessarily required restricting the amount of countries studied.

\begin{figure}
\centering
\includegraphics[width=0.9\textwidth]{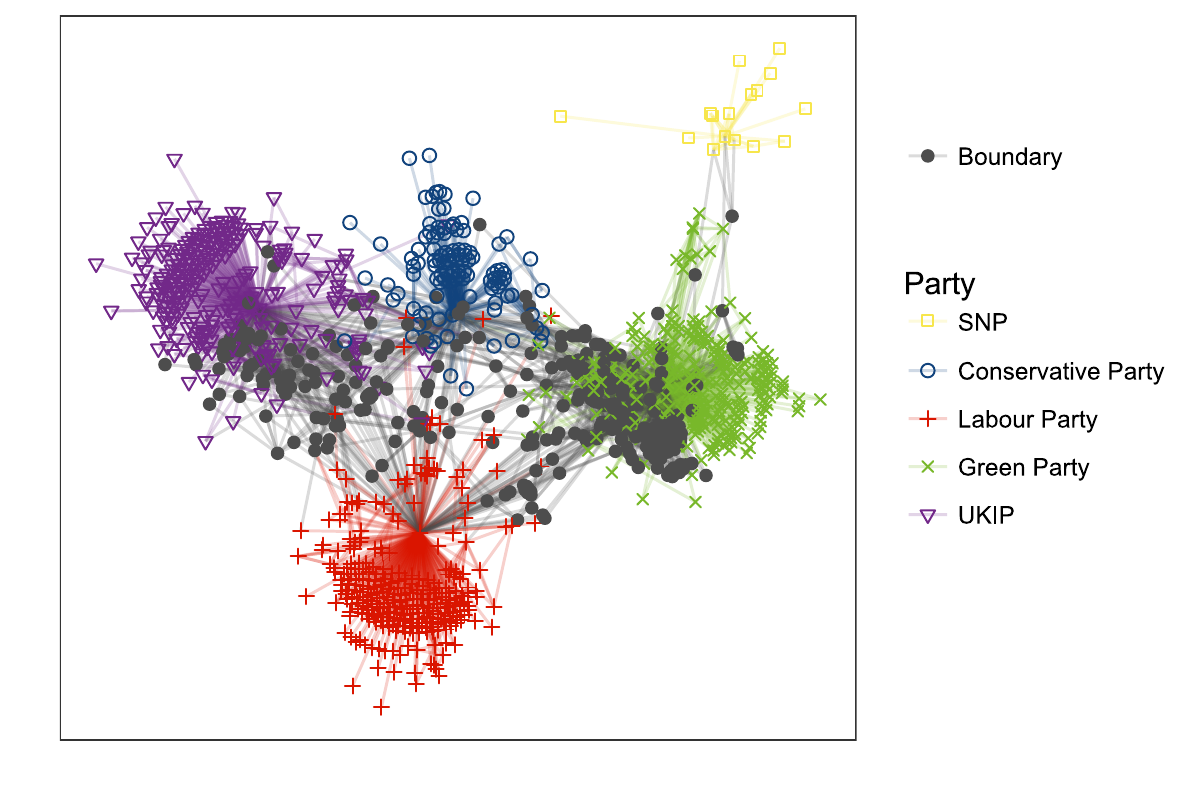}
\caption{\label{fig:uk-one}Structure of observed discussion network in the UK (one day of data only)}
\end{figure}

Figure \ref{fig:uk-one} presents an illustration of the structure of overall discussion networks collected for each country, and how different groups interact within them. The visualization contains one day of data from the UK (the day before the election): choosing a restricted time window for this illustrative figure was necessary as the full data has so many nodes as to make effective visualisation difficult (the actual analysis below is however based on data from a longer time period). The nodes on the network are positioned using the Fruchterman-Reingold algorithm \citep{Fruchterman1991}, a method which is designed to highlight division and group structure within networks. Nodes which communicated solely within a party grouping are given the colour of that party. Nodes which connected to two or more such groupings, which will be referred to as ``boundary'' nodes \citep{Guerra2013}, are coloured in dark grey. 

The graphic illustrates several points. Many of the nodes are almost isolated, connecting to just one of the central party nodes. These represent people who tweeted just once on the day mentioning an individual party in their tweet, and who themselves were never mentioned by anyone else. However there are also significant patterns of conversation within party groupings (indicated by links between nodes of the same colour). Finally, and most importantly for our purposes, there are significant communication patterns \textit{across} groups, indicated by the grey nodes and grey links which form the boundary between groups. 

The data collection window ran from the \nth{11} of May to the \nth{10} of June. This period was chosen because it provided a sample of data from both before and after the elections took place, which makes it possible to test the sensitivity of results to the election itself (i.e. it makes it possible to see whether patterns observed before the election persist after it). The precise day of the election varies between countries, and indeed some elections took place over several days: however all of them were held between the \nth{22} and the \nth{25} of May. In total 1,426,620 tweets were observed which made reference to at least one of the parties in the list. The Twitter streaming API sets a maximum upper limit on the amount of Tweets which can be collected through it. However, the volume of Tweets collected here was comparatively small compared to that limit, meaning that a complete set of data was collected for the time period in question. 

Usage of Twitter varied widely between parties. Some parties had accounts which were mentioned hundreds of thousands of times, whilst others were never mentioned at all. There was also significant variation between countries: in some political spaces all major parties were represented on Twitter, whilst in others just one or two parties were represented. In total 90 parties had enough data to form part of the analysis, from 23 different countries (in this case, ``enough'' data means being mentioned at least once on Twitter, and coming from a country which has at least one other party which also fulfilled those criteria, allowing a pair of parties to be formed). 

The measurement of the extent to which a pair of party groups was fragmented took place in two stages. First, a directed network was created for each pair of political parties within each country (although potentially interesting, the fragmentation of pairs of party groups from different countries arguably raises a different set of theoretical questions, and hence was not studied here). In total this resulted in 164 pairs. This pair level network, which is a subset of the overall discussion network for that country, consists of all individuals who mentioned at least one of the two party accounts for that pair of political parties. These individuals are the nodes in the network, and directed edges between the nodes are created if one individual mentioned another during the data collection period in the form described in Table \ref{edge-create} above. The edges are weighted, with the weight of the edge being the number of times one individual has mentioned another (though the sensitivity of the results to this weighting is tested below).

The network formed for each pair is then separated into three parts: two groups of ``internal'' nodes, which communicated solely with one of the party accounts or with others who also communicated solely with those accounts; and one group of ``boundary'' nodes, which is a group of nodes who communicated with at least one member of both of the sets of internal nodes. Following Guerra et al. (\citeyear{Guerra2013}), the measure of fragmentation developed below is based on the size of the boundary and its patterns of communication, relative to those found within the sets of internal nodes: as the boundary grows and becomes more active, relative to internal communication, fragmentation itself is held to drop. Controlling for the activity levels of internal nodes (rather than just measuring the size of the boundary) is useful in this case because it offers a measure of fragmentation which is essentially independent of size, and hence allows for comparison pairs of groups of different sizes, which is of course key in this context considering the wide variation in Twitter usage.  

The exact metric used to measure fragmentation between a pair of groups, $F$, is given by the following formula:

\begin{equation}
F = \frac{B_{e} - I_{e}}{B_{e} + I_{e}}
\end{equation}

Where $B_{e}$ represents the weighted sum of edges going from the group of boundary nodes to the two groups of internal nodes, and $I_{e}$ represents the weighted sum of edges within the two groups of internal nodes. This formula is conceptually similar to Krackhardt and Stern’s $E-I$ ratio \citep{Krackhardt1988}, a measure which seeks to capture the extent to which a given group is more or less externally facing in its patterns of communications, and which has previously used in studies of online political fragmentation \citep{Hargittai2008}. It is also similar to the measure $P$ developed by Guerra et al. to measure polarization (\citeyear{Guerra2013}). The key difference between $F$ and the measures $E-I$ and $P$ is that $F$ does not require a hard assignment of nodes within the boundary to belong to one group only. Rather, it simply measures patterns of communication and connection of networks surrounding two Twitter accounts of theoretical interest. This avoids the requirement having to develop a separate means of assignment of nodes into individual groups.

While use of $F$ fits well to the task at hand, it is nevertheless worth acknowledging its limitations. It is a ``naive'' measure of fragmentation: it contains no information about whether the individuals are supporters or not of a given political group, nor what type of messages they are sending to each other (which could be supportive, or highly critical, or anything in between). $F$ simply measures the strength of connection between a pair of Twitter handles of interest. This is likely to understate the true level of fragmentation within a pair: as even if some communication is observed, it might be highly critical (whilst in-group communication might be highly positive). It could be the case that conversation networks elicited using (for example) only positive messages might show very different patterns. However, this is again a necessary sacrifice: robustly assessing the content of messages sent in a wide variety of different European languages is beyond the scope of the present paper. Furthermore, as the aim is to explain variation between party pairs, a metric which has the same potential for error for all party pairs is still acceptable. 

It is worth highlighting as well that the measure $F$ also differs from the other major means of measuring fragmentation employed in the literature thus far, which involves the use of some form of automatic community detection algorithm to classify nodes in the network into communities, often supplemented with a graphical approach which lays out the network using an algorithm which aids community identification \citep[for an overview of community detection techniques see][]{Fortunato2010}. The observed communities can then be manually labelled based on qualitative investigation of the participants or discussion themes within them \citep{Aragon2013,Conover2011,Himelboim2013}; a modularity score can also be calculated \citep{Newman2006} to provide an indication of the extent to which the detected community structure ``fits'' the network. However, this approach is not appropriate for this study, because there is no guarantee that the communities which emerge will map onto ideologies measured at the party level. This would make it difficult to then assess the relationship between ideology and the metrics of fragmentation, which are based on these clusters. 

\begin{figure}
\centering
\includegraphics[width=0.9\textwidth]{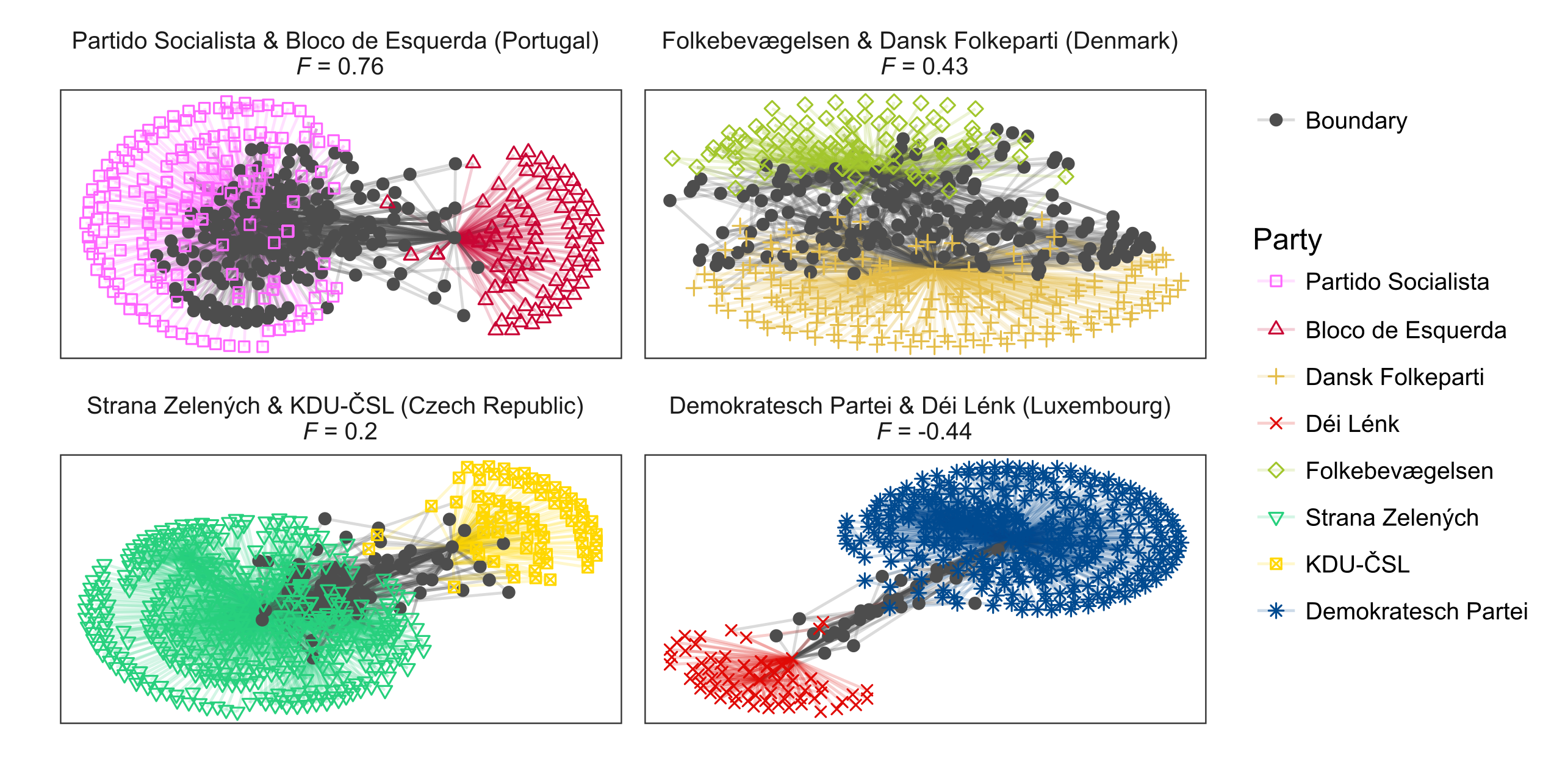}
\caption{\label{fig:pair-ex}Party pairs at different levels of $F$}
\end{figure}

Figure \ref{fig:pair-ex} provides an illustrative visualization of how $F$ relates to the overall network structure for a pair of groups. Four different party pairs are visualised at a range of different levels of $F$. Pairs with relatively small absolute amounts of nodes were selected to make the visualisations easier to interpret. The figure illustrates how the $F$ score decreases as the size of the boundary decreases relative to the amount of non-boundary nodes. At higher levels of $F$ the boundary starts to dominate the overall graphic. The metric, in other words, seems to capture the basic intuition behind fragmentation. However, the figure also highlights a potential concern, which is that even though $F$ controls for the absolute number of nodes within the two groups (and is hence comparable across pairs of groups with different sizes), it seems that it may be sensitive to the ratio of sizes of the two sub groups within the network (for example, the very low $F$ score in the figure relates to a pair comprising a relatively large and relatively small group). The sensitivity of models to this factor is hence tested below.  

The main independent variable in the study is the ideology of political groupings. In order to operationalize the idea of ideology, parties in the dataset were classified on a left-right scale \citep[see e.g.][]{Castles1984}. Left wing parties tend to favour policies relating to economic redistribution and equality; whilst right-wing parties favour policies relating to individual liberty and the free market. Parties on the extremes of either scale favour more radical and far-reaching visions of these policies. Such scales are certainly simplifications of the complexity of political life, and many more complicated classifications exist (in the particular case of European Parliament elections, classification on a pro/anti European integration axis is also common). Yet they are also widely used in both popular discourse and academic research as a way of understanding politics and classifying political groups, and represent the simplest way of classifying a wide variety of groups. The particular classification for each party were taken from the Parlgov dataset \citep{Doring2012}, which places each party on a 10 point scale ranging from 0 (extreme left) to 10 (extreme right). A score of 5 would represent the center ground, though no party scored exactly 5 (the ideology scores assigned to each party can be found in the appendix). 

Positioning parties on an ideological scale allowed calculation of the three main independent variables which relate to each of one of the three hypotheses. First, the ideological distance between parties could be measured, which is the absolute value of the difference between their two ideology scores. Second, parties were coded as belonging to either the left or right of the political scale. Finally, the ``extremism'' of a party’s ideology could be measured, which is simply the absolute value of the distance of their ideology score from the centre ground score of 5. This variable, obviously, takes a value between 0 and 5 at the level of the individual party; the total extremism within a pair of parties can therefore range from 0 to 10.  

Three control variables were used. The size of a political grouping was given by the amount of votes in achieved in the 2014 election, measured as a percentage of total votes. Having these sizes allows specification of the size difference between groups, with an expectation that a greater different in size would result in more fragmentation. The political status of the party in question was also recorded (i.e. whether it was incumbent in government or in opposition). This allows a specification of the relationship between parties: whether they are both incumbent in government, both in opposition, or whether it is an incumbent-opposition pair. Finally, the ratio of tweets observed for the pair of parties was measured, which is simply the amount of tweets observed for the smallest party divided by the amount observed for the largest party. This variable runs from 0 to 1, and controls for potential effects on the $F$ score caused by imbalances in size between the two groupings in question, as described above in the discussion of Figure \ref{fig:pair-ex}.    

\section{Analysis}
Initial descriptive statistics on the data collected are presented in table \ref{desc-stats}. As expected there is wide variation in the levels of fragmentation ($F$) between different groups, with the observed data covering almost the full range of possible values (though the data is skewed to the left)\footnote{The skewed nature of the dependent variable suggested a transformation might improve model fit, hence alternative versions of the models presented in table \ref{tab:results-one} were tried with a cubed dependent variable, which offered a better approximation of normality. However the results were the same, hence the non-transformed variable has been preserved here to facilitate the interpretation of the results.}. There is also considerable variety in the absolute sizes of pairs in terms of number of participants (nodes), with some having less than 100 and some having tens of thousands. Of course, as $F$ scales for the absolute size of the group, this isn't in itself a problem. However the smaller groups do present a concern in that, when the number of contributions is small in absolute terms, the extent to which the observed level of fragmentation is affected by individual nodes and edges is increased: hence there is greater potential for small group $F$ scores to be unreliable. This point will be revisited in the analytical section. 

\begin{table}[h]
\centering
\caption{Descriptive statistics for pairs of parties}
\label{desc-stats}
\begin{tabularx}{.8\textwidth}{lXXXX}
\hline
\hline
Numerical Variables & Min & Max & Mean & SD \\ 
\hline
Observed Nodes & 54 & 175,299 & 14,349 & 28,930 \\
Observed Edges & 109 & 1,088,381 & 76,180 & 150,751\\ 
$F$ & -0.96 & 1.00 & 0.61 & 0.48 \\
Ideological Distance & 0.00 & 8.3 & 2.95 & 1.91 \\
Extremism & 0.00 & 8.3 & 4.17 & 1.42 \\
Size Difference & 0.00 & 34.90 & 9.68 & 8.77 \\
\end{tabularx}
\begin{tabularx}{.8\textwidth}{llXX}
\hline
Categorical variables & & N & \% \\
\hline
Incumbency Status & Incumbent - Incumbent Pair & 25 & 15\% \\
 & Incumbent - Opposition Pair & 84 & 51\% \\
 & Opposition - Opposition Pair & 55 & 34\% \\
Left-Right Status & Left - Right Pair & 72 & 44\%  \\
& Left-Left / Right-Right Pair & 92 & 56\% \\
\hline
Total observations & 164 \\
\hline\hline
\end{tabularx}
\end{table}

The analysis makes use of a series of hierarchical linear (multilevel) models\footnote{Network metrics were calculated using the \texttt{igraph} package in Python. Hierarchical linear models were estimated using the \texttt{lme4} package in R. A replication dataset and accompanying R code file is available in the appendix.}. A hierarchical model is appropriate in this circumstance considering that the observations are drawn from different countries and that variation in the overall extent of fragmentation might well be expected between different political systems. Model fit was assessed through the use of marginal and conditional R$^{2}$ values \citep{Nakagawa2013}. The marginal R$^{2}$ gives a measure of the amount of variance explained by the fixed effects alone while the conditional R$^{2}$ gives the total amount of variance explained by the full model: comparing the two hence also reveals the amount of variance explained by the country level differences. 

The main results are presented in Table \ref{tab:results-one}. This table contains two sets of models, one addressing the effect of ideological distance between parties (models 1.1 to 1.3), and one addressing the effect of extremism (models 2.1 to 2.3). These variables are addressed in separate models because, as might be  expected, they are highly correlated. 

Model 1.1 investigates the effect of two key variables of interest, ideological distance and mismatching left right pairs, on the fragmentation index. Model 1.2 includes a number of theoretically relevant control variables, whilst model 1.3 limits the observations to only party pairs with at least 1,000 nodes, to control for potential effects caused by very small party pairs. The results for ideological distance in model 1.1 go in the expected direction and are statistically significant: as parties get further apart in ideological space, communication between them decreases. This result becomes insignificant in model 1.2, but is fully significant again in model 1.3. Hence overall there is reasonable support for the idea that increasing ideological distance between parties causes increasing fragmentation (Hypothesis 1). The term for a mismatching left-right pair is also significant in model 1.3 (and on the borderline for model 1.1). Interestingly, it is also positive, indicating that pairs from different sides of the left-right divide communicate more than those on the same side. Hence, for example, a pair formed of an extreme left and centre-left party would communicate less than a pair formed of a centre-left and centre-right party. 

One way of explaining this counter-intuitive finding is that centrist parties communicate more than extremist parties. This idea is tested in models 2.1 to 2.3, which look at the impact of extremism rather than ideological distance. The extremism term (which is the sum of both extremism scores for the two parties in the pair) is statistically significant in all models. This shows that as one or both parties in a party pair tend towards the ideological extremes, communication patterns between them drop and fragmentation increases. This provides strong support for Hypothesis 3 which specified that increasing extremism ought to cause more fragmentation. The term for a left-right mismatch is, meanwhile, statistically insignificant in these models, meaning that overall there is little support for Hypothesis 2 (which specified that parties from different sides of the left right divide ought to interact less). 

It is worth briefly commenting on the control variables in models 1.1 – 2.3. There is some evidence that an increasing size difference between parties leads to weaker patterns of communication, suggesting that larger parties tend to ignore smaller ones. There was only weak evidence that incumbency or opposition status made a difference. Meanwhile, the term for the tweet ratio was significant in all models, confirming the sensitivity of $F$ to size imbalances between party pairs. Is is also worth highlighting the R$^{2}$ scores. Conditional R$^{2}$ scores are quite high in all cases, whilst marginal R$^{2}$ scores are quite low. This indicates a large amount of variation at the country level.  Marginal R$^{2}$ rises to its highest levels in models 1.3 and 2.3. These models operate on the subset of party pairs with at least 1,000 nodes. As suspected above, this may indicate that $F$ scores for very small party pairs are more unreliable.

\setlength{\tabcolsep}{0.1pt}

\begin{table}[!htbp] \centering 
  \caption{Hierarchical Linear Models explaining Fragmentation ($F$) for party pairs. All numerical variables are standardized.} 
  \label{tab:results-one} 
\begin{tabular}{>{\raggedright}p{0.2\linewidth}D{.}{.}{-1} D{.}{.}{-1} D{.}{.}{-1} D{.}{.}{-1} D{.}{.}{-1} D{.}{.}{-1} } 
\\[-1.8ex]\hline 
\hline \\[-1.8ex] 
 & \multicolumn{6}{c}{\textit{Dependent variable: $F$ score for party pair}} \\ 
\cline{2-7} 
\\[-1.8ex] & \multicolumn{1}{c}{(1.1)} & \multicolumn{1}{c}{(1.2)} & \multicolumn{1}{c}{(1.3)} & \multicolumn{1}{c}{(2.1)} & \multicolumn{1}{c}{(2.2)} & \multicolumn{1}{c}{(2.3)}\\ 
\hline \\[-1.8ex] 
  Ideology & -0.27^{*} & -0.13 & -0.31^{**} &  &  &  \\ 
  Extremism &  &  &  & -0.22^{***} & -0.17^{**} & -0.23^{***} \\ 
  Left-Right Pair & 0.33^{\dagger} & 0.18 & 0.55^{**} & 0.01 & 0.01 & 0.11 \\ 
  Size Difference &  & -0.05 & -0.16^{*} &  & -0.064 & -0.17^{*} \\ 
  I-O Pair$^{\ddag}$ &  & -0.23 & -0.21 &  & -0.173 & -0.17 \\ 
  O-O Pair$^{\ddag}$ &  & -0.32 & -0.49^{*} &  & -0.19 & -0.39 \\ 
  Tweet Ratio & & 0.33^{***} & 0.29^{***} & & 0.32^{***} & 0.28^{***} \\
 \hline \\[-1.8ex] 
Observations & \multicolumn{1}{c}{164} & \multicolumn{1}{c}{164} & \multicolumn{1}{c}{128} & \multicolumn{1}{c}{164} & \multicolumn{1}{c}{164} & \multicolumn{1}{c}{128} \\ 
Marginal R$^{2}$ & 0.02 & 0.11 & 0.25 & 0.04 & 0.12 & 0.26 \\
Conditional R$^{2}$ & 0.60 & 0.74 & 0.62 & 0.61 & 0.74 & 0.63 \\
\hline 
\hline \\[-1.8ex] 
\textit{Notes:}  & \multicolumn{6}{r}{$^{\dagger}$p$<$0.1; $^{*}$p$<$0.05; $^{**}$p$<$0.01; $^{***}$p$<$0.001} \\ 
& \multicolumn{6}{>{\raggedleft}p{0.75\linewidth}}{$^{\ddag}$When compared to a pair of incumbent parties, the effect of being either an incumbent - opposition [I-O] pair or an opposition - opposition [O-O] pair} \\
\end{tabular} 
\end{table}

One potential objection to the claims above is that, while the models suggest that the presence of extremism within a party pair is important, they do not reveal whether it is extremist parties themselves who communicate less, or whether it is centrists who refuse to communicate with extremists. This question is explored in Table \ref{tab:results-two}. This table presents two further hierarchical linear models. Here the level of observation is a party in a pair (rather than a party pair), hence there are 328 observations in the full model. These models make use of a different dependent variable which measures a level of fragmentation specific to the individual party within a pair. Hence the equation for $F$ is rewritten as:

\begin{equation}
F_{p} = \frac{B_{e} - I_{ep}}{B_{e} + I_{ep}}
\end{equation}

Where $B_{e}$ continues to represent connections from the boundary outwards but $I_{ep}$ represents internal connections within the given party (rather than both parties in the pair). 

Making use of this new dependent variable, model 3.1 looks at all parties, while model 3.2 is limited to those with at least 100 nodes. The findings in both models are essentially identical and support the conclusions drawn above: more extreme parties communicate less with other parties (relative to their own internal patterns of communication). There is also a clear finding for party size, with larger parties, in terms of vote share, communicating less than smaller ones. 

\begin{table}[!htbp] \centering 
  \caption{Hierarchical Linear Models explaining Fragmentation ($F_{p}$) for single parties within a pair. All numerical variables are standardized.} 
  \label{tab:results-two} 
\begin{tabular}{lD{.}{.}{-2}D{.}{.}{-2}} 
\\[-1.8ex]\hline 
\hline \\[-1.8ex] 
 & \multicolumn{2}{c}{\textit{Dependent variable: $F_{p}$ for a given party}} \\ 
\cline{2-3} 
\\[-1.8ex] & \multicolumn{1}{c}{(3.1)} & \multicolumn{1}{c}{(3.2)}\\ 
\hline \\[-1.8ex] 
  Extremism & -0.12^{*} & -0.14^{*} \\ 
  Right-wing & -0.01 & -0.10 \\ 
  Party Size & -0.16^{**} & -0.25^{***} \\ 
  Incumbent & -0.19 & -0.23^{\dagger} \\ 
 \hline \\[-1.8ex] 
Observations & \multicolumn{1}{c}{328} & \multicolumn{1}{c}{242} \\ 
Marginal R$^{2}$ & 0.04 & 0.09 \\
Conditional R$^{2}$ & 0.12 & 0.25 \\
\hline 
\hline \\[-1.8ex] 
\textit{Note:}  & \multicolumn{2}{r}{$^{\dagger}$p$<$0.1; $^{*}$p$<$0.05; $^{**}$p$<$0.01; $^{***}$p$<$0.001} \\
\end{tabular} 
\end{table}

A further potential objection to these findings is that it is not clear how sensitive they are to alternative means of eliciting conversation networks from Twitter. The networks employed make use of both ``mention'' and ``retweet'' data: yet mentions and retweets have been shown to exhibit different network structures in previous research \citep{Conover2011}. Furthermore, the data collection window encompasses both the before and after election period, yet conversation patterns have been shown to be different before and after this type of critical juncture \citep{Garcia2015}. Finally, making use of ``weighted'' networks creates the potential for small numbers of very active users, who may send a lot of messages and hence create very strong connections, to have a disproportionate influence on the results \citep{Bermudez2016}

To test robustness to these alternative means of eliciting networks, five final models are specified, which can be found in Table \ref{tab:results-robust}. Each one is identical to model 2.2 above, yet employs an alternative means of eliciting the conversation network: model 4.1 uses only mention data, 4.2 uses only retweet data, 4.3 uses the before election period, 4.4 uses the after election period and 4.5 uses an unweighted network instead of a weighted one.

The results are robust to all of these potential different specifications with the exception of the retweet data, where the term still points in the same direction but is no longer statistically significant (p = 0.15). This supports the idea that mention and retweet data may produce quite different network structures (though it is also worth noting there were far fewer retweets observed than mentions, which could also explain the discrepancy found here). Interestingly, the effect of extremism is also stronger after the election period than it is before it, indicating perhaps that extremist parties are more connected to mainstream political discourse at election time. Overall, however, these further models offer good support for the idea that the results are robust to different network specifications. 

\begin{table}[!htbp] \centering 
  \caption{Further Models. All numerical variables are standardized.} 
  \label{tab:results-robust} 
\begin{tabular}{>{\raggedright}p{0.2\linewidth}D{.}{.}{-2} D{.}{.}{-2} D{.}{.}{-2} D{.}{.}{-2} D{.}{.}{-2}} 
\\[-1.8ex]\hline 
\hline \\[-1.8ex] 
 & \multicolumn{5}{c}{\textit{Dependent variable: $F$ score for party pair}} \\ 
\cline{2-6} 
 & \multicolumn{1}{c}{Mentions} & \multicolumn{1}{c}{Retweets} & \multicolumn{1}{c}{Pre-elec.} & \multicolumn{1}{c}{Post-elec.} & \multicolumn{1}{c}{Unweighted} \\ 
\\[-1.8ex] & \multicolumn{1}{c}{(4.1)} & \multicolumn{1}{c}{(4.2)} & \multicolumn{1}{c}{(4.3)} & \multicolumn{1}{c}{(4.4)} & \multicolumn{1}{c}{(4.5)} \\ 
\hline \\[-1.8ex] 
  Extremism & -0.08^{*} & -0.07 & -0.09^{*} & -0.14^{***} & -0.13^{**} \\ 
  Left-Right Pair & 0.03 & -0.05 & -0.01 & 0.04 & -0.03 \\ 
  Size Difference & -0.03 & -0.06 & -0.07^{\dagger} & -0.01 & -0.08^{\dagger}\\ 
  I-O Pair$^{\ddag}$ & -0.16^{\dagger} & -0.19 & -0.09 & -0.28^{**} & -0.32^{**} \\ 
  O-O Pair$^{\ddag}$ & -0.23^{\dagger} & -0.21 & -0.08 & -0.33^{*} & -0.29^{*} \\ 
  Tweet Ratio & 0.13^{***} & 0.07 & 0.14^{***} & 0.17^{***} & 0.07^{*} \\
 \hline \\[-1.8ex] 
Observations & \multicolumn{1}{c}{164} & \multicolumn{1}{c}{163} & \multicolumn{1}{c}{164} & \multicolumn{1}{c}{164} & \multicolumn{1}{c}{164}\\ 
Marginal R$^{2}$ & 0.12 & 0.07 & 0.10 & 0.20 & 0.21 \\
Conditional R$^{2}$ & 0.61 & 0.37 & 0.68 & 0.60 & 0.33 \\
\hline 
\hline \\[-1.8ex] 
\textit{Notes:}  & \multicolumn{5}{r}{$^{\dagger}$p$<$0.1; $^{*}$p$<$0.05; $^{**}$p$<$0.01; $^{***}$p$<$0.001} \\
& \multicolumn{5}{>{\raggedleft}p{0.65\linewidth}}{$^{\ddag}$When compared to a pair of incumbent parties, the effect of being either an incumbent - opposition [I-O] pair or an opposition - opposition [O-O] pair} \\
\end{tabular} 
\end{table} 

\section{Discussion}

This article has sought to provide a first large scale analytical treatment of the reasons behind the emergence of echo chambers, defined as the widely observed phenomenon of online networks self-organising into groups along ideological lines. Based on a novel dataset drawn from Twitter, and working at the party-pair level, it has shown that fragmentation varies within networks, with some pairs of groups communicating more than others. It found good evidence supporting the idea that the ideological distance between groups plays a role in explaining this variation, with particular evidence that extremist parties communicate less, such that centrist parties (even from different sides of the left-right divide) were more likely to communicate than a centrist and extremist party from the same side of the left-right divide. It also found no evidence that left and right wing parties have different communication patterns, contradicting a range of pieces of previous research. These results were robust to a variety of alternative model specifications. 

These findings have implications for the theory of online political fragmentation and indeed online discussion more generally. Several points can be made. One of the major worries advanced by theorists of political fragmentation has concerned exposure to alternative viewpoints, with worries that the development of online echo chambers will lead to people hearing their own views repeated again and again. The evidence presented here, by contrast, shows that lots of communication does occur between different ideologies, especially across the left-right divide but within the centre ground. In fact, the real area of separation appears to occur with people who hold extreme ideologies, who become separated both from people of other viewpoints and even people who hold more moderate versions of their viewpoint. This may indicate that the most important factor is, as Stroud has suggested (\citeyear{Stroud2010}), the certainty with which people hold beliefs, rather than general ideological differences between individuals. 

Another factor concerns some of the other determinants of fragmentation. The evidence presented here adds weight to the idea that the position of a political party within the political system changes the way they interact with technology: larger parties seemed to generally communicate less with other groups. It is intriguing to find that parties which are more politically successful offline are also typically more disconnected online, and it is significant because it shows that online fragmentation is not purely a result of decisions made by individuals online; the offline context has an impact. However the evidence also undermines the idea that left and right wing parties have fundamentally different communication patterns, and hence that qualitative differences between these two types of ideology generate more or less fragmentation. In the future, single country studies which report such differences should be treated with scepticism. 

It is worth concluding by highlighting areas which the present paper was unable to address, thus indicating potential directions for further research. Three points stand out. First, no separation was made in the paper between different types of Twitter user, in particular no discrimination was made between people who use Twitter for professional political purposes (such as party members, politicians and journalists) and the general public. But we might expect these political professionals to show very different communication patterns. Second, as the data in the paper is a snapshot, it is unable to address temporal dynamics, which much of the literature have suggested might play a role in fragmentation (for example, whether groups become more extreme over time). Finally, as already highlighted above, the paper does not address the content of the messages, rather just the structure and volume of communication. As research continues in these areas, we will continue to understand more about what drives the emergence of echo chambers in online discussion networks.  

\printbibliography 

\end{document}